# PECULIAR AGNs FROM THE INTEGRAL AND RXTE DATA

**E. Fedorova[1], PhD, A. Vasylenko[2], PhD, V.I. Zhdanov[1], Dr.Sci., Prof.,**
**[1]Astronomical Observatory of National Taras Shevchenko University of Kyiv**
**[2]Main Astronomical Observatory of the NAS of Ukraine**

*We analysed how the active galactic nucleus (AGN) X-ray primary continuum depends on AGN activity in radio, using the data of RXTE Spectral & Timing Database (Rossi X-ray Timing Explorer) and INTEGRAL (INTErnational Gamma-Ray Astrophysics Laboratory). Our aim is to test the relation between spectral shapes of these AGNs and the prediction of the "spin-paradigm" model of the AGN "central engine". We have found that for the major part of radio-quiet (RQ) AGNs the value of high-energy exponential cut-off in primary spectrum appears to be significantly higher than 100 keV and thus follows the "spin-paradigm" predictions. In the same time, near 25% of radio-loud (RL) AGNs demonstrate the high-energy cut-off at the energies above 150 keV, contradicting the "spin paradigm". We have composed a sample of "peculiar" 69 RQ and 10 RL AGNs that seem to contradict to the "spin paradigm" for further individual studies of these objects.*

**Introduction.** One of the most prominent physical differences between AGNs of various classes can be related with jet activity and RL/RQ dichotomy. There are several schemes explaining physical (non-geometrical) differences in the AGN structure, and one of the most well-known of them is often referred to as the "spin-paradigm". This model associates the jet activity with high values of the central super massive black hole (SMBH) spin as well as with the direction of rotation of the accretion disk (AD) around the black hole [1]. Namely, high values of the SMBH spin (a>0.75) in prograde system (i.e. when the directions of the SMBH spin and the angular velocity of accretion disk rotation are the same) and retrograde systems (i.e. systems with SMBH spin antiparallel to the angular velocity of AD) with the spin of SMBH a<-0.1 correspond to powerful jet and radio emission in RL AGN with the radio loudness parameter $R = L_{vradio} / L_{vopt} > 100$ (here $L_{vradio}$ and $L_{vopt}$ are monochromatic luminosities at 5GHz and in V band, correspondingly [2]). The comparably low values of SMBH spin (-0.1<a<0.5) corresponds to RQ AGNs with low values of radio loudness parameter (R<100, [2]) and weak or no jets[3].

Following this model, in RQ AGNs fit into standard Shakura-Synyaev AD model [4] (i.e. geometrically thin, radiatively efficient, steady state disk with zero viscosity at the sonic point) with almost radial infall inside the innermost stable orbit; whereas the disks in RL AGNs are "torqued" (there is nonzero magnetic viscosity at the sonic point) magnetized disks described by Agol and Krolik [5]. In such systems the spin energy of central SMBH can dissipate through the sonic point into the disk due to magnetic reconnection between the inner part of AD and SMBH horizon. One of the consequences of this model is that the innermost part of AD in RL AGN is disrupted and smeared away by powerful centrifugally-driven jet outflows, leading to the low values of the exponential cutoff in the spectrum of the primary emission (below 100 keV), whereas the high values or absence of the high-energy cut-off are prescribed to RQ AGN spectrum. However, some objects were found which seem to contradict this paradigm [6-9]. That is why it is interesting to compare the high-energy exponential cut-off with the RL/RQ characteristics of AGNs.

In the present paper we obtain main parameters of the hard X-ray spectra (0.5-250 keV) of 79 non-blazar Seyfert galaxies from RXTE spectra & timing database to find the objects with peculiar spectral shape. Additionally, we consider for this aim the objects of INTEGRAL sample we worked out earlier in [10].

In Sections 2 we describe the samples of AGNs we considered, and describe the data and models, and present the best fit individual model parameters of the AGNs of RXTE sample.

In Section 3, we show the resulting subsample of peculiar AGNs from both RXTE and INTEGRAL samples. Finally, in the last section we discuss our results and draw out conclusions.

**Sample, data and fitting.** We compiled a sample of AGNs from the RXTE AGN spectra and timing database. After excluding blazars and bad datasets, the final sample consists of 79 AGNs, including 10 RL and 69 RQ AGNs. We also included in our consideration the INTEGRAL sample of 95 AGNs we treated earlier in [10].

To perform our spectral fitting, we use the XSPEC 12.8 package of HEASOFT (High Energy Astrophysical SOFTware) software provided by NASA [11].

For the primary spectrum of AGN we used the standard power law with an exponential cut-off at high energies $A_{PL}(E) = KE^{-\Gamma} \exp(-E/E_c)$ (*cutoffpl*), where $\Gamma$ is the photon index and $E_c$ is the cut-off energy. This primary emission is reflected from a neutral or ionized material of an accretion disc or gas-dust torus; to model it we used *pexrav* model [12]. This model describes a power-law emission partially reflected by an infinite flat slab of neutral or ionized medium. To take into account the proper absorption in the investigated object, described by the formula $A_{abs}(E) = \exp\{-N_H \sigma[E(1+z)]\}$ and modeled by *zphabs* model in XSPEC. Galactic absorption (in the Milky Way) is also included in all the spectral models, following the data by Kalberla et al.[13] and using the model *phabs* (the same as *zphabs* with the redshift z=0). Thus the model in XSPEC for S1-S1.5 AGNs schematically looks like:

**phabs($N_{HGal}$)*zphabs*pexrav( $\Gamma$, $E_c$, R).**

describing the emission of an AGN "central engine" both direct and reflection fractions (pexrav) with the same low level of absorption, and for S1.7 - S2 AGNs:

**phabs($N_{HGal}$)*(zphabs*cutoffpl( $\Gamma$, $E_c$, R)+pexrav( $\Gamma$, $E_c$, R)),**

describing the sum of strongly obscured direct emission of the "central engine" (*cutoffpl*) and slightly absorbed reflected one (*pexrav*).

For each galaxy of the sample, we obtained such spectral parameters as the power-law index, relative reflection parameter $R$, cut-off energy $E_c$ and intrinsic absorption value. The best-fit parameters are presented in Table 1, together with the errors, lower and upper limits at 90% confidence level. Below we perform also a correlation analysis of the main spectral parameters, namely "photon index - cut-off energy". The peculiar AGNs addressing the "spin-paradigm" are boldfaced and underlined in the Table 1. Objects with the best-fit value of the high-energy cut-off contradicting the "spin-paradigm" predictions, but with error's level too high to determine this finally (i.e. "candidates to the peculiar subsample"), are underlined only.

Table 1. Best fit parameters of AGNs of RXTE sample.

| Source | Class | $\Gamma$ | $E_c$, keV | R | $N_H$, $10^{22}$ cm$^{-2}$ |
|---|---|---|---|---|---|
| Mkn 335 | RQ NL S1 | 2.60±0.05 | >170 | 4.6±2.7 | 2.3±1.4 |
| **RHS 03 (RBS 78)** | RQ S1 | 1.37±0.12 | $8^{+18}_{-3}$ | $2.2^{+1.5}_{-1.3}$ | <0.95 |
| Mkn 348 | RQ S2 | 1.77±0.07 | >100 | <7.0 | 24±4 |
| **PG 0052+251** | RQ S1.2 | 1.75±0.04 | $73^{+42}_{-18}$ | <0.45 | <1.0 |
| **TONS 180** | RQ NL S1 | 2.09±0.15 | 34±21 | <0.3 | <0.5 |
| Fairall 9 | RQ S1 | 1.79±0.03 | >15 | 3.8±0.3 | <1.0 |
| **NGC 526A** | RQ S1.9 | 1.25±0.05 | 10±1 | 4.2±0.7 | 4100±400 |
| **Mkn 590** | RQ S1.2 | $1.59^{+0.14}_{-0.13}$ | $18^{+9}_{-4}$ | $2.7^{+1.8}_{-1.3}$ | <1.0 |
| NGC 1052 | RL S2 | 1.97±0.05 | $36^{+9}_{-6}$ | $24^{+6}_{-5}$ | 22±3 |
| NGC 1068 | RQ S2 | 1.62±0.12 | >1400 | <1.16 | >930 |
| 4U 0241+61 | RQ S1.2 | 1.94±0.05 | $160^{+189}_{-74}$ | 1.21±0.26 | 3.35±0.33 |
| RHS 17 | RQ S1 | 2.50±0.11 | >70 | 18±5 | <1.0 |
| 4U 0241+611 | RQ S1.2 | 1.92±0.06 | 76±64 | 1.5±0.4 | 3.2±0.4 |
| NGC 1386 | RQ S2 | 3.5±1.3 | >3 | <47 | $2.0^{+0.7}_{-0.8}$ |
| **3C 111** | RL S1 | 1.88±0.01 | >1500 | 0.16±0.05 | 2.7±0.2 |
| **3C 120** | RL S1 | 1.92±0.01 | 230±40 | 0.38±0.06 | 0.45±0.33 |
| **IRAS 04575-7537** | RQ S2 | 2.28±0.04 | 15±1 | 10±2 | $24^{+26}_{-10}$ |
| **Ark 120** | RQ S2 | 2.07±0.04 | 26±4 | 3.5±0.4 | 0.8±0.2 |
| **Pictor A** | RQ S1 | 1.35±0.03 | 13±1 | 1.0±0.3 | <1.0 |
| E253-G3 | RQ S2 | $3.18^{+0.8}_{-1.9}$ | >15 | <35 | $151^{+42}_{-37}$ |
| NGC 2110 | RQ S2 | 2.09±0.05 | >500 | 1.1±0.5 | 24±2 |
| **MCG+8-11-11** | RQ S1 | 1.40±0.04 | 11±1 | 4.8±0.6 | <1.0 |
| PKS 0558-504 | RQ NL S1 | 2.17±0.03 | >289 | <0.22 | <1.0 |
| Mkn 3 | RQ S2 | 1.71±0.06 | >300 | <3.3 | $8.8^{+3.4}_{-3.3}$ |
| **Mkn 79** | RQ S1.2 | 1.95±0.03 | $19^{+30}_{-4}$ | 8.7±0.9 | <0.9 |
| PG 0804+761 | RQ S1 | 2.12±0.23 | >20 | 2.4±1.2 | <2.4 |
| **PKS 0921-213** | RL S1 | 1.70±0.06 | $132^{+1700}_{-66}$ | $0.50^{+0.42}_{-0.39}$ | <0.81 |
| **Mkn 110** | RQ NL S1.5 | 1.79±0.02 | $52^{+25}_{-13}$ | 0.98±0.01 | 0.81±0.20 |
| **NGC 2992** | RQ S2 | $2.00^{+0.10}_{-0.09}$ | $16^{+3}_{-2}$ | $12.6^{+4.1}_{-7.8}$ | $15^{+14}_{-13}$ |
| MCG-5-23-16 | RQ S2 | 2.21±0.03 | >500 | 0.7±0.2 | 33±2 |
| NGC 3227 | RQ S1.5 | 1.20±0.03 | >1300 | 1.9±0.3 | $23^{+5}_{-6}$ |
| NGC 3281 | RQ S2 | 2.6±0.4 | >30 | <4.0 | $300^{+9}_{-7}$ |
| NGC 3516 | RQ S1.5 | 2.14±0.06 | >209 | 2.9±0.5 | 21±3 |
| **PG 1116+215** | RQ S1 | 2.35±0.11 | 18±4 | 23±7 | <0.5 |
| NGC 3783 | RQ S1 | 2.00±0.02 | 900±400 | 0.95±0.13 | <0.5 |
| NGC 3998 | RQ S1 | 2.00±0.06 | $106^{+690}_{-50}$ | <0.39 | <0.5 |
| NGC 4051 | RQ NL S1.5 | 2.59±0.01 | $397^{+540}_{-146}$ | 7.1±0.3 | 1.5±0.2 |
| PG 1202+281 | RQ S1 | 2.2±0.5 | >10 | 0.88±0.09 | <0.5 |
| NGC 4151 | RQ S1.5 | 1.53±0.06 | $112^{+32}_{-8}$ | 9.8±3.3 | 9.8±0.3 |
| **PG 1211+143** | RQ NL S1 | 2.41±0.05 | $27^{+11}_{-4}$ | 3.9±1.2 | 9.8±0.3 |
| **Mkn 766** | RQ NL S1 | 2.12±0.09 | $23^{+7}_{-4}$ | >3.5 | <0.5 |
| **NGC 4258** | RQ S2 | 1.34±0.17 | $11^{+4}_{-2}$ | 1.9±1.5 | <0.5 |
| NGC 4388 | RQ S2 | 2.15±0.19 | >500 | 5.6±4.7 | 6.4±0.4 |
| **TON 1542 (Mkn 771, Ark 374, RBS 1125)** | RQ S1 | 1.8±0.6 | $10^{+21}_{-4}$ | <1 | <0.5 |
| NGC 4507 | RQ S2 | 1.77±0.12 | >500 | 0.57±0.05 | 6.4±0.6 |
| NGC 4593 | RQ S1 | 1.82±0.01 | $1130^{+2291}_{-471}$ | 4.2±0.8 | <0.5 |
| NGC 4945 | RQ S2 | 1.22±0.09 | 143±44 | 4.2±0.8 | 3.2±0.2 |
| **Centaurus A (NGC 5128)** | RL S2 | 1.86±0.02 | >1000 | <0.0002 | 12±1 |
| MCG-6-30-15 | RQ NL S1.2 | 2.42±0.01 | >2008 | 3.1±0.1 | 2.7±0.2 |
| **IRAS 13349+2438** | RQ S1 | 1.7±0.8 | $8^{+20}_{-3}$ | <22 | |
| IC 4329A | RQ S1.2 | 2.05±0.05 | >300 | 0.83±0.09 | 15±3 |
| **Mkn 279** | RQ S1.5 | 1.7±0.1 | 12±1 | 4.4±1.3 | <0.5 |
| Circinus | RQ S2 | 3.25±0.05 | >2000 | >1000 | 54±2 |
| NGC 5506 | RQ S1.9 | 2.3±0.1 | >800 | 4±1 | 13±3 |
| NGC 5548 | RQ S1 | 1.98±0.6 | 287±118 | 0.9±0.2 | 14±2 |
| PG 1416-129 | RQ S1 | 1.4±0.2 | $503^{+3505}_{-405}$ | <4.5 | <0.5 |
| PG 1440+356 (Mkn 478) | RQ NL S1 | $3.5^{+0.6}_{-2.5}$ | >5 | <38.3 | <0.5 |
| MCG-2-40-4 | RQ S2 | 1.8±0.7 | >7 | <6.8 | $80^{+24}_{-16}$ |
| NGC 6251 | RQ S2 | 2.27±0.15 | >60 | 1.3±0.5 | 680±120 |
| NGC 6240 | RQ S2 | 1.96±0.28 | >605 | 5.1±0.6 | 130±20 |
| NGC 6300 | RQ S2 | 2.41±0.09 | $354^{+360}_{-250}$ | >46 | $80^{+13}_{-10}$ |
| PDS 456 | RQ S1 | 3.8±0.2 | >200 | 42±25 | <0.5 |
| **3C 382** | RL S1 | 1.97±0.01 | >117 | 0.39±0.07 | 1.89±0.23 |
| **IRAS 1825-5926** | RQ S2 | 1.9±0.6 | 10±6 | <8.8 | 5.2±1.8 |
| ESO 103-G35 | RL S2 | 2.33±0.02 | >70 | >40 | 36±1 |
| **3C 390.3** | RL S1 | 1.79±0.04 | $126^{+73}_{-37}$ | 0.25±0.11 | 1440±250 |
| **3C 405 (Cygnus A)** | RL S2 | 2.81±0.13 | >1000 | <1.7 | 46±1 |
| RHS 56 | RQ NL S1 | 2.4±0.4 | >41 | 4.5±2.1 | <0.5 |
| Mkn 509 | RQ S1.2 | 1.91±0.01 | $466^{+211}_{-114}$ | 0.51±0.05 | 0.68±0.15 |
| IC 5063 | RQ S2 | 2.22±0.03 | >290 | 28±12 | 44±2 |
| NGC 7172 | RQ S2 | 2.04±0.02 | >540 | 3.7±1.1 | 25±12 |
| NGC 7213 | RL S1.5 | 1.59±0.08 | 10±3 | 4.9±2.0 | <0.01 |
| NGC 7314 | RQ S1.9 | 2.25±0.12 | >600 | 1.9±0.5 | 27±7 |
| **Ark 564** | RQ NL S1 | 1.70±0.07 | $27^{+7}_{-4}$ | 0.6±0.2 | 12.2±0.3 |
| MR 2251-178 | RQ S1.5 | 1.78±0.01 | $225^{+67}_{-43}$ | <0.03 | 1.0±0.2 |
| NGC 7469 | RQ S1.2 | 2.1±0.9 | >800 | 0.76±0.25 | 3.0±0.4 |
| MCG-2-58-22 (Mkn 926) | RQ S1.5 | 1.90±0.07 | $303^{+375}_{-50}$ | 0.39±0.15 | 3.4±0.3 |
| NGC 7582 | RQ S2 | 2.35±0.09 | >728 | 120±80 | 50±9 |
| **RHS 61** | RQ S1 | 1.81±1.11 | $8^{+40}_{-4}$ | <14 | <1.0 |

The $\Gamma$ vs. $E_c$ diagram for both the objects of RXTE and INTEGRAL samples is shown on the Fig.1. Correlations for parameters for the total set of the data are shown in the Table 2.

Table 2. Correlation coefficients between spectral parameters for the AGNs of RXTE sample.

| Correlation, parameters -> Object class ↓ | $\Gamma$, $E_c$ | $\Gamma$, R | $E_c$, R |
|---|---|---|---|
| RQ | 0.08 | 0.28 | 0.196 |
| RL | 0.58 | -0.09 | 0.26. |
| all | 0.19 | 0.26 | 0.14 |

As we can see we have small correlations between photon index and reflection parameter, as well as between the cut-off energy and reflection parameter, however, we see that the correlation between photon index and cut-off energy is considerable.

At the same time, the mean values of these spectral parameters over the subsamples of RL and RQ AGNs do not differ within the error limits: $\Gamma$=1.91±0.13 and $E_c = 296^{+690}_{-37}$ keV for RQ AGNs, and $\Gamma$=1.73±0.08 and $E_c = 390^{+1000}_{-25}$ keV for RL ones.

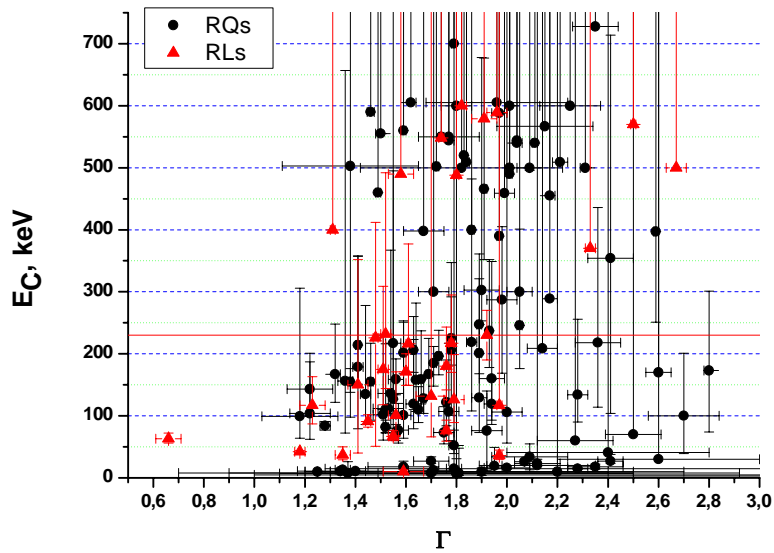

Figure 1. Photon index vs. cut-off energy diagram for the AGNs of RXTE and INTEGRAL samples.

**Peculiar AGNs and spin-paradigm.** Analysis performed had revealed the subsample of AGNs deviating from the "spin-paradigm" prescriptions. The "mainstream" of RQ AGNs really follows this pattern, and only 25 objects from 150 (15% of total number of RQ objects) demonstrate the exponential high-energy cut-off below or near 100 keV. The situation with RL objects is opposite: only 17 of total 35 objects (48%) have high-energy cut-offs below 100 keV in their spectra. Additionally to the results of the Table.1, the peculiar AGNs found in the INTEGRAL sample are shown in the Table.3 (for these objects the same spectral model was applied in [10] to describe hard X-ray continuum emission above 3 keV). Here the "confirmed" peculiar AGNs with high enough accuracy level of parameters determining are also boldfaced. These sources should be recommended for further more detailed investigations of the full set of data available on them to understand exactly the reason of their contradiction with spin paradigm. Presumably, such high percentage of "spin-paradigm-deviating" AGNs among the RL party can be caused by jet contamination of a primary spectrum of the "central engine".

Table 3. Peculiar AGNs from RXTE and INTEGRAL samples. The dash in the last column ($E_c$) denotes the absence of high-energy exponential cut-off in the spectrum of an object (cut-off has not been detected).

| Name | Class | $\Gamma$ | R | $E_c$, keV |
|---|---|---|---|---|
| IGR J18027-1455 | RQ S1 | $1.52\pm0.03$ | $0.63^{+0.68}_{-0.38}$ | $110^{+106}_{-38}$ |
| WKK 1263 | RQ S1 | $1.63\pm0.03$ | $1.51^{+0.66}_{-0.60}$ | $119^{+104}_{-40}$ |
| **GRS 1734-292** | RQ S1 | $1.52\pm0.01$ | $0.38^{+0.15}_{-0.19}$ | $82^{+11}_{-9}$ |
| IGR J16482-3036 | RQ S1 | $1.59^{+0.04}_{-0.02}$ | $1.49^{+1.09}_{-0.68}$ | $101^{+150}_{-37}$ |
| IGR J17488-3253 | RQ S1 | $1.44\pm0.02$ | $0.18\pm0.18$ | $135^{+143}_{-49}$ |
| SWIFT J1038.8-4942 | RQ S1 | $1.51\pm0.04$ | $1.25\pm0.38$ | $102^{+128}_{-41}$ |
| LEDA 090443 | RQ S1 | $2.28\pm0.04$ | $6.26^{+1.36}_{-1.17}$ | $134^{+122}_{-44}$ |
| **IGR J07597-3842** | RQ S1 | $1.56\pm0.01$ | $0.61^{+0.35}_{-0.33}$ | $71^{+15}_{-11}$ |
| NGC 4593 | RQ S1 | $1.76\pm0.01$ | $0.69\pm0.11$ | $122^{+68}_{-38}$ |
| 2E 1853.7+1534 | RQ S1 | $1.89\pm0.03$ | $2.24^{+0.93}_{-0.81}$ | $129^{+232}_{-56}$ |
| 1A 1343-60 | RQ S1 | $1.65\pm0.02$ | $0.58\pm0.19$ | $110^{+37}_{-21}$ |
| **NGC 4151** | RQ S1 | $1.28\pm0.02$ | $0.05\pm0.03$ | $84\pm7$ |
| **M 87** | RL S1 | $2.50\pm0.01$ | $0$ | - |
| **IGR J13109-5552** | RL S1 | $1.58\pm0.05$ | $1.58^{+1.35}_{-1.15}$ | - |
| **3C 382** | RL S1 | $1.76\pm0.01$ | $0.68\pm0.19$ | $180^{+63}_{-38}$ |
| **3C 111** | RL S1 | $1.60\pm0.01$ | $0.29\pm0.11$ | $171^{+29}_{-22}$ |
| **3C 390.3** | RL S1 | $1.78\pm0.01$ | $1.28\pm0.17$ | $217^{+78}_{-47}$ |
| **Pictor A** | RL S1 | $1.80\pm0.01$ | $0.99\pm0.23$ | - |
| **3C 120** | RL S1 | $1.74\pm0.01$ | $0.25^{+17.79}_{-0.09}$ | - |
| **QSO B0241+62** | RL S1 | $1.61\pm0.01$ | $1.05\pm0.13$ | $216^{+161}_{-67}$ |
| **S5 2116+81** | RL S1 | $1.96\pm0.04$ | $2.37^{+1.15}_{-0.99}$ | - |
| **WKK 6471** | RL S1 | $1.91\pm0.05$ | $0$ | - |
| MCG-01-24-012 | RQ S2 | $1.77\pm0.04$ | $1.88^{+1.22}_{-1.02}$ | $107^{+110}_{-47}$ |
| ESO 103-35 | RQ S2 | $1.94\pm0.03$ | $1.89^{+0.82}_{-0.69}$ | $119^{+40}_{-26}$ |
| PGC 037894 | RQ S2 | $1.54^{+0.06}_{-0.04}$ | $1.42^{+1.53}_{-1.04}$ | $136^{+231}_{-56}$ |
| WKK 0560 | RQ S2 | $1.22^{+0.10}_{-0.07}$ | $1.44^{+2.97}_{-1.73}$ | $104^{+97}_{-42}$ |
| NGC 4138 | RQ S2 | $1.36\pm0.03$ | $0.09^{+0.72}_{-0.01}$ | $156^{+501}_{-84}$ |
| NGC 1194[1] | RQ S2 | $1.18\pm0.15$ | $1.15\pm0.89$ | $99^{+207}_{-35}$ |
| ESO 506-27 | RQ S2 | $1.46\pm0.01$ | $0.73\pm0.34$ | $155\pm62$ |
| IGR J20187+4041 | RQ S2 | $1.57\pm0.04$ | $1.48^{+1.31}_{-1.07}$ | $77^{+43}_{-21}$ |
| IC 4518A | RQ S2 | $1.55\pm0.01$ | $2.62\pm0.61$ | $127^{+663}_{-66}$ |
| NGC 4992 | RQ S2 | $1.41\pm0.02$ | $0$ | $179^{+180}_{-100}$ |
| NGC 6240 | RQ S2 | $2.70^{+0.11}_{-0.16}$ | $0$ | $100^{+101}_{-61}$ |
| NGC 1052 | RL S2 | $1.41\pm0.04$ | $0.10^{+0.34}_{-0.05}$ | $150^{+202}_{-110}$ |
| Mrk 6 | RL S2 | $1.48\pm0.01$ | $0.52\pm0.13$ | $226^{+186}_{-175}$ |
| **3C 405** | RL S2 | $1.51\pm0.02$ | $0$ | $175^{+113}_{-57}$ |
| **NGC 1275** | RL S2 | $2.67\pm0.04$ | $0.59\pm0.39$ | - |
| **NGC 5128** | RL S2 | $1.82\pm0.02$ | $0$ | - |
| **NGC 5252** | RL S2 | $1.31\pm0.01$ | $0$ | - |

**Conclusions.** We have performed the spectral modelling of the RXTE data of 79 non-blazar AGNs from RXTE spectra & timing database, to test the correspondence between their spectral properties and the pattern predicted by the "spin-paradigm" model of AGN "central engine". We also used to this purpose the original sample of 95 AGNs observed by INTEGRAL mission which we have investigated earlier in [10].

As a result we have revealed 25 RQ and 17 RL AGNs demonstrating a behaviour contradicting the "spin-paradigm" prescribing low values of high-energy exponential cut-off in RL AGN spectra and higher ones for RQ AGNs. These objects have to be studied in details to clear out the reasons of such contradiction. For instance, the most probable reasons of the great percentage of peculiar RL object can be in fact "fake" caused by the X-ray contamination from the jet. However, in the same time we note, especially for RQ objects, that the percentage of peculiar objects can be a bit higher due to uncorrelated variability in different ranges of energies [15]: a part of objects with a variable spectral shape like NGC 4388 or NGC 4945 [7, 14], can be missed in an investigation like that of the present paper.

### Acknowledgements

This research has made use of data and software obtained through the RXTE and INTEGRAL science data centers (ESA); provided by the NASA/Goddard Space Flight Center. The work has been supported by the scientific program "Astronomy and Space Physics" of Taras Shevchenko National University of Kyiv.

**References:**
1. Garofalo D., Evans D.A., Sambruna R.M.//MNRAS. -2010. - Vol.406. - P.975.
2. Sikora M., Stawarz L., Lasota J.-P.// http://www.slac.stanford.edu/cgi-wrap/getdoc/slac-pub-12284.pdf
3. Garofalo D.//Adv. Astron. – 2013. - ID 213105.
4. Shakura N. I., Sunyaev R. A.//A&A. -1973. - Vol.24. - P.337.
5. Agol E., Krolik J.H.//Astron.J. – 2000. - Vol. 528. - p.161-170.
6. De Rosa A., Bassani L., Ubertini P. et al.// A&A. – 2008. - Vol.483. - p.749.
7. Fedorova E., Beckmann V., Neronov A., Soldi S.//MNRAS. – 2011. - Vol.417. - p.1140.
8. Soldi S., Beckmann V., Gehrels N., De Jong S., Lubinski P.//2011, in: proc.of the Workshop "Narrow-Line Seyfert 1 Galaxies and Their Place in the Universe". ArXiv:1105.5993.
9. Soldi S., Beckmann V.,Bassani L., Courvoisier T.J.-L. et al. //A&A. – 2005. - Vol.444. - p.431.
10. Vasylenko A., Zhdanov V., Fedorova E.//Astroph. Space Sci. – 2015. – Vol.360. – P.71.
11. Arnaud K., Gordon C., Dorman B. An X-Ray Spectral Fitting Package //
https://heasarc.gsfc.nasa.gov/xanadu/xspec/manual/manual.html
12. Magdziarz, P., Zdziarski, A. A. //MNRAS. – 1995. – Vol.273. – P. 837.
13. Kalberla P. M. W., Burton W. B., Hartmann D., Arnal E. M., Bajaja E., Morras R., Poppel W. G. L.//A&A. – 2005. – Vol.440. – P.775.
14. Fedorova E., Zhdanov V.I.//Kinemat.Phys.Celest.Bodies. – 2016. - Vol. 32, is.4. - p.172-180.
15. Chesnok N.G., Sergeev S.G., Vavilova I.B.// Kinemat. Phys. Celest. Bodies. - 2009. - Vol. 25, No. 2. - P. 107 – 113.